\documentclass[prd,superscriptaddress,twocolumn,amsfonts,amssymb,amsmath]{revtex4}

\usepackage{graphicx}
\usepackage{dcolumn}
\usepackage{bm}
\usepackage{natbib}
\usepackage{color}

\newcommand{\C}[1]{{\mathcal #1}}

\newcommand{\beq}{\begin{equation}}
\newcommand{\eeq}{\end{equation}}
\newcommand{\bea}{\begin{eqnarray}}
\newcommand{\eea}{\end{eqnarray}}
\newcommand{\nn}{\nonumber}

\newcommand{\half}{\frac 12}
\newcommand{\third}{\frac 13}
\newcommand{\quarter}{\frac 14}

\newcommand{\Slash}[1]{{\ooalign{\hfil#1\hfil\crcr\raise.167ex\hbox{/}}}}

\begin{document}

\preprint{arXiv:yymm.nnnn}

\title{Supersymmetric standard model inflation in the Planck era}

\author{Masato Arai}
\affiliation{
Institute of Experimental and Applied Physics,
Czech Technical University in Prague, 
Horsk\' a 3a/22, 128 00 Prague 2, Czech Republic}
\author{Shinsuke Kawai}
\affiliation{
Institute for the Early Universe (IEU),
11-1 Daehyun-dong, Seodaemun-gu, Seoul 120-750, Korea} 
\affiliation{Department of Physics, 
Sungkyunkwan University,
Suwon 440-746, Korea}
\author{Nobuchika Okada}
\affiliation{
Department of Physics and Astronomy, 
University of Alabama, 
Tuscaloosa, AL35487, USA} 

\date{\today}

\begin{abstract}

We propose a cosmological inflationary scenario based on the supergravity-embedded
Standard Model supplemented by the right-handed neutrinos.
We show that with an appropriate 
K\"{a}hler potential the $L$-$H_u$ direction gives rise to successful inflation that is similar to the
recently proposed gravitationally coupled Higgs inflation model but is free from the unitarity 
problem.
The mass scale $M_R$ of the right-handed neutrinos is subject to the seesaw relation and 
the present 2-$\sigma$ constraint from the WMAP7-BAO-$H_0$ data sets
its lower bound $M_R\gtrsim$ 1 TeV.
Generation of the baryon asymmetry is naturally implemented in this model.
We expect that within a few years new observational data from the Planck satellite will clearly
discriminate this model from other existing inflationary models arising from the same Lagrangian,
and possibly yield stringent constraints on $M_R$.
\end{abstract}

\maketitle

\section{Introduction}
Today observational cosmology is a precision science.
Cosmological inflation, which is supported by all observational data, is now an indispensable
theoretical ingredient not only in astrophysics but also in particle phenomenology.
A remaining mystery of this otherwise extremely successful paradigm is embedding into
a particle theory model. 
By virtue of Occam's razor, a plausible 
possibility may be that the fields responsible for
cosmological inflation (inflatons) are those already included in the Standard Model (SM), 
or its (not too large) extension.
The recently proposed SM Higgs inflation model \cite{HiggsInflation}
is an interesting idea to test this possibility. 
This model is attractive due to its minimalistic nature and the remarkable 
agreement with the present day observational data.
It also relates the dynamics of inflation with the electroweak scale physics, making a prediction
on the SM Higgs mass from the cosmological microwave background (CMB) data.
A rather unfavourable feature of this type of model is that it requires extremely large 
nonminimal coupling to gravity, which could lead to violation of the unitarity bound
\cite{unitarity}.
The model also suffers from the hierarchy problem, which may be cured by 
supersymmetrisation 
\cite{Einhorn:2009bh,FKLMP,Arai:2011nq}. 
See \cite{Pallis:2011ps} for related models.

Certainly, there are more traditional ways of embedding inflation into supersymmetric SMs.
It has been known for a while that the flat directions in the minimal supersymmetric Standard 
Model (MSSM), lifted by soft supersymmetry breaking terms and other effects, 
can serve as inflatons
(reviewed in \cite{Enqvist:2003gh}; more recent developments include
\cite{ATI}).
Another type of embedding is into a supersymmetric SM with right-handed neutrinos
\cite{MSYY},
in which one of the right-handed sneutrinos is identified as the inflaton.
These models are phenomenologically well motivated; 
the hierarchy problem is solved by supersymmetry, 
and the models with the right-handed neutrinos are furthermore 
consistent with the small but nonzero neutrino masses indicated by neutrino oscillation.

In this paper we present a new scenario of inflation, inspired by these
developments.
Our model has the following features:
(i) the scenario is based on the simplest supersymmetric extension of the SM that includes
the right-handed neutrinos,
naturally explaining the small neutrino masses through the seesaw mechanism \cite{seesaw};
(ii) the problem associated with the large nonminimal coupling that afflicts the SM Higgs inflation
is alleviated;
(iii) the CMB data gives predictions on the mass scale of the right-handed neutrinos through the
seesaw relation;
(iv) leptogenesis is naturally implemented;
(v) the predicted cosmological parameters fit well in the present day observational constraint, 
and
(vi) the model can be tested by the upcoming observational data from the Planck satellite. 
We discuss construction of the model and describe these features below.

\section{The supersymmetric seesaw model}
Our model is based on the MSSM extended with the right-handed neutrinos, with
the $R$-parity preserving superpotential
\beq
W=W_{\rm MSSM}+\half M_{R} N_R^cN_R^c+y_D N_R^cLH_u,
\label{eqn:W_SSM}
\eeq
where $N_R$ is the right-handed neutrino superfield (having odd $R$-parity), 
$M_{R}$ the mass parameter for $N_R$, and 
\beq
\hspace{-2pt} W_{\rm MSSM}=\mu H_uH_d+y_u u^cQH_u+y_d d^cQH_d+y_e e^cLH_d,
\eeq
is the MSSM part. 
Here, $Q$, $u$, $d$, $L$, $e$, $H_u$, $H_d$ are the MSSM superfields, $\mu$ the MSSM $\mu$-parameter, 
and $y_D$, $y_u$, $y_d$, $y_e$ the Yukawa couplings 
(the family indices are suppressed). 
As noted in \cite{Einhorn:2009bh}, successful nonminimally coupled Higgs inflation
requires at least an extra field besides those in the MSSM.
Our crucial observation here is that the model (\ref{eqn:W_SSM}) is already such an extension,
with the $L$-$H_u$ direction playing the r\^{o}le of inflaton.
During inflation $Q$, $u$, $d$, $e$, $H_d$ do not play any part and we shall disregard them.
Parametrising the D-flat direction along $L$-$H_u$ as
\beq
L=\frac{1}{\sqrt 2}\left(\begin{array}{c} \varphi\\ 0 \end{array}\right),
\quad
H_u=\frac{1}{\sqrt 2}\left(\begin{array}{c} 0\\ \varphi\end{array}\right),
\eeq
the superpotential becomes
\beq
W=\half M_{R}N_R^cN_R^c+\half y_D N_R^c\varphi^2.
\label{eqn:W}
\eeq
We assume supergravity embedding and choose 
\beq
\Phi
=1-\third \left(|N_R^c|^2+|\varphi|^2\right)
+\quarter\gamma\left(\varphi^2+{\rm c.c.}\right)+\third\zeta|N_R^c|^4,\label{eqn:Kahler}
\eeq
with $\gamma$ and $\zeta$ real parameters.
The K\"{a}hler potential in the superconformal framework is $K=-3 \Phi$.
We have included an $R$-parity violating term. 
For brevity's sake, we shall set the reduced Planck scale $M_{\rm P}=2.44\times 10^{18}$ GeV
to be unity, take $y_D$ to be real and consider only one generation below.

We introduce real scalar fields $\chi$, $N$, $\alpha_1$, $\alpha_2$ by
$\varphi=\frac{1}{\sqrt 2}\chi e^{i\alpha_1}$,
$N_R^c= Ne^{i\alpha_2}$.
It can be checked that the scalar potential is stable along the real axes of $\varphi$ and
$N_R^c$; we thus assume $\alpha_1=\alpha_2=0$ below.
The scalar-gravity part of the Lagrangian in the Jordan frame reads 
(cf. \cite{FKLMP})
\beq
\hspace{-3pt}
{\C L}_{\rm J}=\sqrt{-g_{\rm J}}\left[\half\Phi R_{\rm J}
-\half g_{\rm J}^{\mu\nu}\partial_\mu\chi\partial_\nu\chi
-\kappa g_{\rm J}^{\mu\nu}\partial_\mu N\partial_\nu N
-V_{\rm J}\right],
\eeq
where 
\beq
\hspace{-3pt}
\Phi=M^2+\xi\chi^2,
\quad
M^2\equiv 1-\third N^2+\frac{\zeta}{3} N^4,
\quad
\xi\equiv\frac{\gamma}{4}-\frac 16.
\eeq
The subscripts J indicate quantities in the Jordan frame, and
$\kappa=1-4\zeta N^2$ is the nontrivial component of the K\"{a}hler metric.
The F-term scalar potential is computed in the standard way
\cite{superconformal}.
In the Jordan frame it reads
\bea
V_{\rm J}&=&\half y_D^2 N^2\chi^2+\frac{(M_{R}N+\quarter y_D\chi^2)^2}{1-4\zeta N^2}\nn\\
&&
\hspace{-18pt}
-\frac{N^2\left\{\half M_{R}N+\frac 34 \gamma y_D \chi^2
-\frac{\zeta N^2(y_D\chi^2+4M_{R}N)}
{2(1-4\zeta N^2)}\right\}^2}{3+\frac{\zeta N^4}{1-4\zeta N^2}
+\frac 34 \gamma\chi^2 (\frac 32 \gamma-1)}.
\eea
The scalar potential in the Einstein frame is $V_{\rm E}=\Phi^{-2} V_{\rm J}$.

In this model the Dirac Yukawa coupling $y_D$ and the right-handed neutrino mass 
$M_{R}$
are 
related by the seesaw relation
\cite{seesaw}
$m_\nu={y_D^2 \langle H_u\rangle^2}/{M_{R}}$,
where $m_\nu$ is the mass scale of the light (left-handed) neutrinos. 
Using the neutrino oscillation data 
$m_\nu^2\approx \Delta m_{32}^2=2.43\times 10^{-3}$ eV${}^2$ \cite{Nakamura:2010zzi}
and the Higgs {\sc vev} at low energy $\langle H_u\rangle \approx 174$ GeV, we find
\beq
y_D=\left(\frac{M_R}{6.14\times 10^{14} \mbox{ GeV}}\right)^{\half}.
\label{eqn:seesaw}
\eeq
This puts an upper bound on $M_R$ since $y_D\lesssim{\C O}(1)$. 


\begin{figure}[h]
\includegraphics[width=50mm]{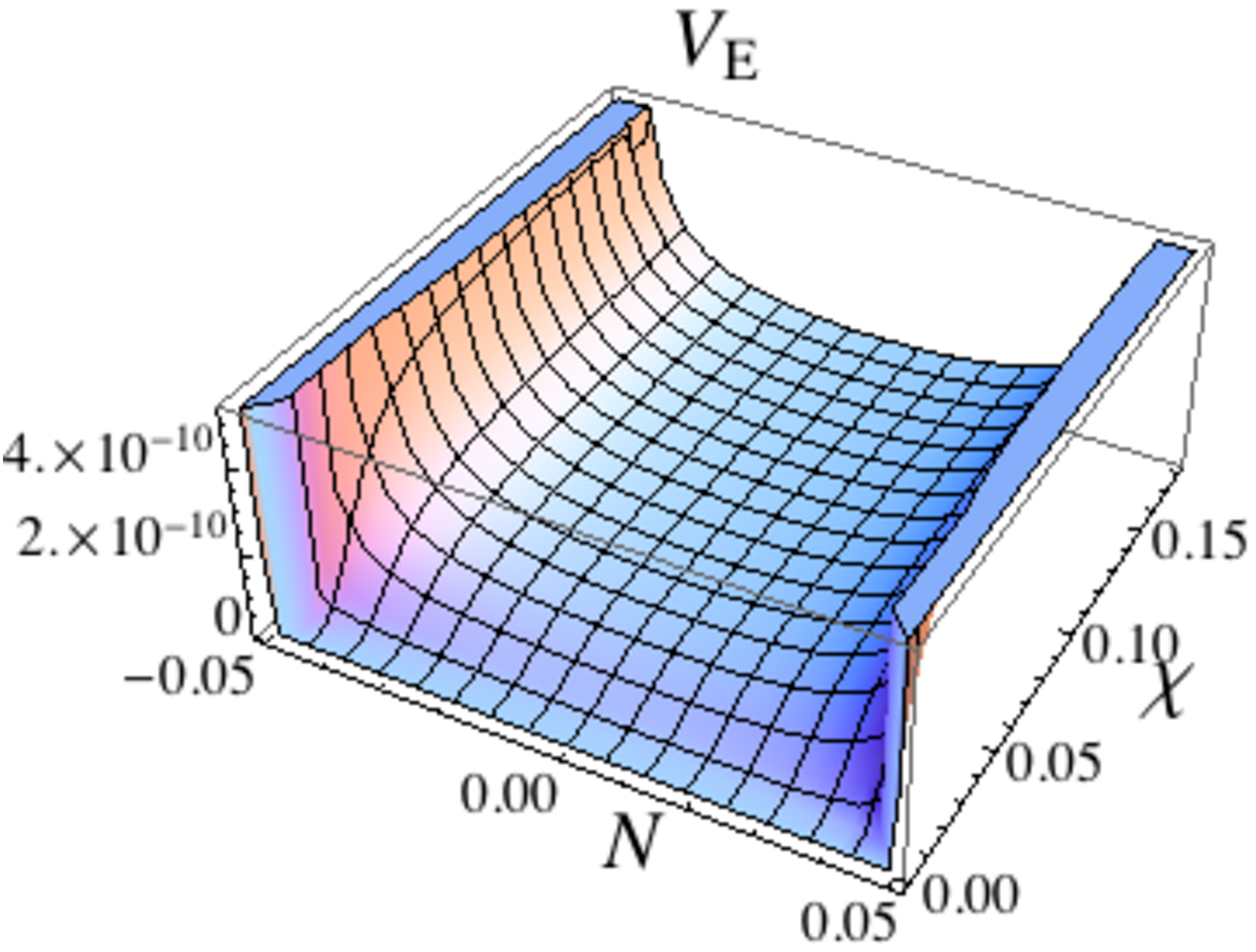}
\includegraphics[width=35mm]{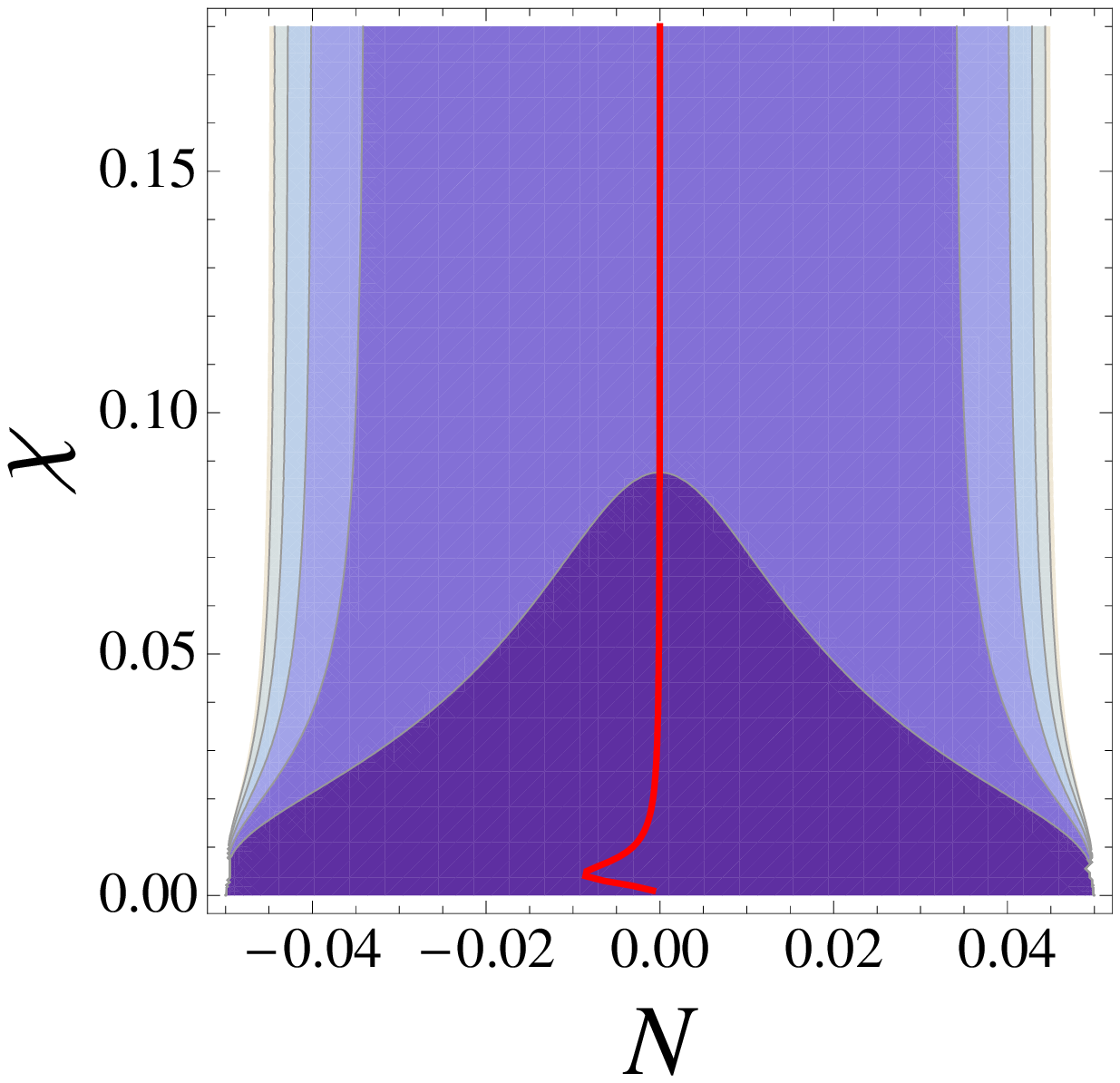}
\caption{\label{fig:potential}
The scalar potential $V_{\rm E}$ in the Einstein frame (left), 
and the inflaton trajectory in the contour plot of the same potential (right). 
The red curve is the inflaton trajectory.
We have chosen $N_e=60$, $M_R=10^{13}$ GeV and $\zeta=100$.}
\end{figure}


For large $y_D$ (and thus large $M_R$) the inflationary model is very similar to the 
next-to-minimal supersymmetric SM 
\cite{Einhorn:2009bh,FKLMP} 
or the supersymmetric grand unified theory model \cite{Arai:2011nq}.
These two-field inflation models in general have nontrivial inflaton trajectories that can
source the isocurvature mode.
While such a scenario is certainly of interest, the analysis is rather involved;
we thus allow the quartic K\"{a}hler term in (\ref{eqn:Kahler}) to control the instability in the
$N$-direction. 
For $M_R=10^{13}$ GeV we find $\zeta=100$ keeps the deviation of $N$ from
$N=0$ negligibly small ($\sqrt{2\kappa}\Delta N/\Delta \chi\lesssim$ 1 \% 
throughout the slow roll of $N_e=60$ e-folds). 
For $M_R\leq 10^{11}$ GeV, $\zeta=1$ is enough.
In Fig.\ref{fig:potential} we show the potential and the inflaton trajectory of our model,
for $M_R=10^{13}$ GeV, $N_e=60$ and $\zeta=100$
(the nonminimal coupling is fixed by CMB as below).
Once the trajectory is stabilised the cosmological parameters are insensitive to the value of
$\zeta$, and as the trajectory is nearly straight the model simplifies to
single field inflation with the inflaton $\chi$.
 The Lagrangian then becomes
\beq
{\C L}_{\rm J}=\sqrt{-g_{\rm J}}\left[\frac{M^2+\xi\chi^2}{2} R_{\rm J}
-\half g_{\rm J}^{\mu\nu}\partial_\mu\chi\partial_\nu\chi-V_{\rm J}\right].
\eeq


\begin{table}[t]
\begin{center}\begin{tabular}{c|c||ccccc}
$N_e$ & $M_R$ (GeV) & $\xi$ & $\chi_*$ & $\chi_k$ & $n_s$ & $r$ \\ \hline
& $10^{13}$ & 2566 & 0.0212 & 0.167 & 0.962 & 0.00419 \\ 
& $10^{11}$ & 257 & 0.0671 & 0.527 & 0.962 & 0.00420 \\
& $10^{9}$ & 25.6 & 0.212 & 1.66 & 0.962 & 0.00422 \\ 
& $10^{6}$ & 0.730 & 1.14 & 8.91 & 0.961 & 0.00515 \\ 
50 & $10^{5}$ & 0.184 & 1.85 & 14.2 & 0.961 & 0.00796 \\ 
& $10^{4}$ & $ 0.0303 $ & 2.79 & 18.9 & 0.960 & 0.0259 \\ 
& $5000$ & $ 0.0152 $ & 3.06 & 19.6 & 0.959 & 0.0448 \\
& $2000$ & $ 4.97\times 10^{-3} $ & 3.31 & 20.1 & 0.955 & 0.103 \\
& $1000$ & $1.33\times 10^{-3}$ & 3.42 & 20.3 & 0.949 & 0.201 \\ 
& $644$ & $ 0 $ & 3.46 & 20.3 & 0.942 & 0.311 \\
\hline
& $10^{13}$ & 3059 & 0.0194 & 0.167 & 0.968 & 0.00296 \\ 
& $10^{11}$ & 306 & 0.0614 & 0.527 & 0.968 & 0.00297 \\
& $10^{9}$ & 30.5 & 0.194 & 1.66 & 0.968 & 0.00298 \\ 
& $10^{6}$ & 0.886 & 1.05 & 8.97 & 0.968 & 0.00352 \\ 
& $10^5$ & $0.232$ & 1.73 & 14.6 & 0.968 & 0.00508 \\ 
60 & $10^4$ & $ 0.0421 $ & 2.63 & 20.1 & 0.967 & 0.0143 \\ 
& $5000$ & $0.0222$ & 2.92 & 21.1 & 0.966 & 0.0237 \\ 
& $2000$ & $8.36\times 10^{-3} $ & 3.22 & 21.8 & 0.964 & 0.0519 \\ 
& $1000$ & $3.28\times 10^{-3}$ & 3.36 & 22.1 & 0.961 & 0.0998 \\ 
& $500$ & $6.48\times 10^{-4}$ & 3.44 & 22.2 & 0.955 & 0.197 \\ 
& $378$& $ 0 $ & 3.46 & 22.2 & 0.951 & 0.260\end{tabular}
\caption{The coupling $\xi$, the inflaton values at the end of the slow roll ($\chi_*$) and at the horizon exit ($\chi_k$), the spectral index $n_s$, and the tensor-to-scalar ratio $r$ for e-folding
$N_e=50$, $60$ and for various values of the right-handed neutrino mass $M_R$.
The coupling $\xi$ is fixed by the amplitude of the curvature perturbation.
We used $\zeta=100$ for $M_R=10^{13}$ GeV and $\zeta=1.0$ for $M_R\leq 10^{11}$ GeV.
The last lines ($N_e=50$, $M_R=644$ GeV and $N_e=60$, $M_R=378$ GeV) 
correspond to the minimally coupled $\lambda\phi^4$ model.
}\label{table:numerical}
\end{center}
\end{table}

\section{Cosmological scenario and the prediction}
Our model provides a cosmological scenario of slow-roll inflation: 
the slow roll parameters $\epsilon$, $\eta$ are small during inflation, and 
inflation terminates when $\epsilon$ or $\eta$ becomes ${\C O}(1)$.
The canonically normalised inflaton field $\hat\chi$ in the Einstein frame is related to $\chi$ by
\beq
d\hat\chi=\frac{\sqrt{M^2+\xi \chi^2+6\xi^2 \chi^2}}{M^2+\xi \chi^2} d\chi,
\eeq
and the slow roll parameters in the Einstein frame are
\beq
\epsilon=\half\left(\frac{1}{V_{\rm E}}\frac{d V_{\rm E}}{d\hat\chi}\right)^2,
\qquad 
\eta=\frac{1}{V_{\rm E}}\frac{d^2 V_{\rm E}}{d\hat\chi^2}.
\eeq
The inflaton value $\chi=\chi_*$ at the end of the slow roll is related to the value
$\chi=\chi_k$ at the horizon exit of the comoving CMB scale $k$, 
through the e-folding number
$N_{e}=\int_{\chi_*}^{\chi_k}d\chi V_{\rm E}({d\hat\chi}/{d\chi})/({d V_{\rm E}}/{d\hat\chi})$.
The potential $V_{\rm E}$ at the horizon exit is constrained by the power spectrum
${\C P}_R=V_{\rm E}/24\pi^2\epsilon$ of the curvature perturbation.
We used the maximum likelihood value $\Delta^2_R(k_0)=2.42\times 10^{-9}$
from the 7-year WMAP data \cite{Komatsu:2010fb}, which is related to the power spectrum 
by $\Delta_R^2(k)=\frac{k^3}{2\pi^2}{\C P}_R(k)$, with the normalisation fixed at 
$k_0=0.002 \mbox{ Mpc}^{-1}$.
Apart from $\zeta$ which was introduced to keep the deviation of the trajectory from $N=0$ small, 
the model contains only two parameters: $\xi$ and $y_D$.
The former is fixed by the curvature perturbation ${\C P}_R$, and the latter is related to the
right-handed neutrino mass $M_R$, through (\ref{eqn:seesaw}). 
Note that there exists a lower bound on $y_D$, 
set by the minimal coupling limit $\xi\rightarrow 0$.
In this limit our model is essentially the chaotic inflation with quartic potential
$V_{\rm E}=\frac{1}{16}y_D^2\chi^4$, with $y_D$ fixed by ${\C P}_R$.
The corresponding value of $M_R$ at $\xi=0$ is 644 GeV for $N_e=50$ and 378 GeV for
$N_e=60$.

For a given value of $M_R$ the scalar spectral index 
$n_s\equiv d\ln{\C P}_R/d\ln k=1-6\epsilon+2\eta$ and the tensor-to-scalar ratio
$r\equiv {\C P}_{\rm gw}/{\C P}_{\rm R}=16\epsilon$ can be computed.
Table \ref{table:numerical} shows these results, 
evaluated for $N_e=50$, $60$ and for several values of $M_R$ between the upper and lower
bounds \footnote{These are the tree-level results. We have checked that the effect of renormalisation is extremely small.}.
We see that $\xi\lesssim {\C O}(1)$ when $M_R\lesssim 10^6$ GeV. 
This shows that in the wide parameter region our model is free from the
dangers 
\cite{unitarity}
arising from the large nonminimal coupling. 
For small $\xi$, instead, a super-Planckian initial value of the inflaton field is inevitable.
This feature is similar to the model studied in \cite{Okada:2010jf}.

After the slow roll the inflaton oscillates around the minimum at $N=\chi=0$,
and decays. 
The effect of nonminimal coupling on the reheating process can be important when
$\xi$ is large and the coupling between the inflaton and the matter field is small
\cite{NonMinReheat}.
In our model, the inflaton couples directly to the SM matter fields and the coupling
$\xi$ does not have to be extremely large; we thus expect the effect of $\xi$ on the reheating to be limited.
The upper limit of the reheating temperature is estimated as $T_{\rm rh}\sim 10^7$ GeV,
assuming the Higgs component decay $\varphi\rightarrow b\bar b$
(the slepton component decay may yield slightly higher temperature
\footnote{Resonance effects can further increase $T_{\rm rh}$. See \cite{Allahverdi:2011aj}.}
).
This is low enough to avoid the gravitino problem.
The generation of the baryon asymmetry is due to the following mechanisms.
If $T_{\rm rh}\gtrsim M_R$, the right-handed (s)neutrinos thermalise, leading to thermal
leptogenesis \cite{Fukugita:1986hr} with the resonant enhancement effects
\cite{ResLep}.
If the reheating temperature is lower $T_{\rm rh}\lesssim M_R$, the mechanism of 
\cite{MSYY,Murayama:1993em} due to the decay of oscillating sneutrinos can be operative;
with $N$ acquiring the {\sc vev} at the end of the slow roll, as shown in Fig.\ref{fig:potential}, 
the coherent oscillation in the direction of $N$ produces lepton numbers. 
Interestingly, this mechanism depends on the inflaton trajectory and thus on $\zeta$.
In addition, the Affleck-Dine mechanism \cite{Affleck:1984fy}
can be operative.
%

\begin{figure}
\includegraphics[width=85mm]{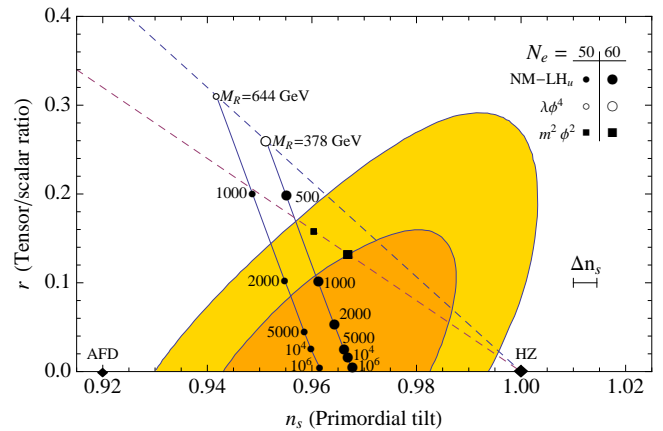}
\caption{\label{fig:WMAP_Plot}
The scalar spectral index $n_s$ and the tensor-to-scalar ratio $r$, with the 68\%
and 95\% confidence level contours from the WMAP7$+$BAO$+H_0$ data \cite{Komatsu:2010fb}.
The prediction of our model (NM-$LH_u$) is indicated by
$\bullet$ with corresponding $M_R$ values.
The predictions of the Harrison-Zel'dovich (HZ), the $\lambda\phi^4$ and $m^2\phi^2$ chaotic
inflation models, as well as the A-term MSSM flat-direction (AFD) inflation models, 
are also shown for comparison. $\Delta n_s$ is the expected Planck accuracy 
\cite{Planck:2006uk}.}
\end{figure}



The prediction of $n_s$ and $r$ in our model is shown in Fig.\ref{fig:WMAP_Plot}, along with
the 68\% and 95\% confidence level contours from the WMAP7$+$BAO$+H_0$ data
\cite{Komatsu:2010fb}.
Also indicated are the predictions of two other inflationary models arising from the same 
Lagrangian (\ref{eqn:W_SSM}), namely the $\tilde N_R$ chaotic inflation model 
\cite{MSYY},
marked with {\small $\blacksquare$},
and the A-term inflation models \cite{ATI} marked with $\blacklozenge$ (AFD).
The former is essentially the standard $m^2\phi^2$ chaotic inflation.
In the latter, the inflaton is $u^cd^cd^c$, $e^cLL$, or $N_R^c LH_u$ direction in the ($N_R$-extended) MSSM, and its typical prediction is very small $r$ and $n_s\approx1-4/N_e$; we used $N_e=50$ (thus $n_s=0.92$) as the e-folding cannot be large ($N_e\lesssim 50$) in such low-scale inflation models.
We see that our model fits well with the present data unless $M_R$ is too small.
The 2-$\sigma$ constraints roughly give $M_R\gtrsim 1$ TeV, depending on the e-folding
number (and thus on the reheating temperature).
In the near future detailed data from the Planck satellite experiments \cite{Planck:2006uk}
will be available, with the expected resolution $\Delta n_s\approx 0.0045$, also indicated in 
Fig.\ref{fig:WMAP_Plot}.
With such high precision the three inflation models arising from the ($N_R$-extended) MSSM
would clearly be discriminated. 
If our model turns out to be the likely scenario, the Planck data would also
constrain the mass scale of the right-handed neutrinos.

\section{Discussion}
%
While the SM of particle theory is the greatest success in the 
twentieth century physics,
it is not a complete theory.
For one thing, 
the neutrino oscillation indicates that the right-handed neutrinos must be 
included.
Also, in order to solve the hierarchy problem and to account for the dark matter in the universe,
some extension, such as supersymmetrisation, is necessary.
In this paper we presented a new scenario of inflation, for which
the right-handed neutrinos, supersymmetry, and the non-minimal coupling are essential.
Note that all of them naturally arise in the supergravity embedding of the SM with the 
right-handed neutrinos.
Not too large nonminimal coupling is also natural as we are dealing with quantum field theory in
curved spacetime.


Our scenario is economical as it explains -- apart from the standard issues 
that are solved by inflation -- small nonvanishing neutrino masses and the origin of the
baryon asymmetry.
The predicted values of $n_s$ and $r$ are consistent with the present 
observation, and can be tested by the Planck satellite data.
What we find particularly interesting is that it constrains the right-handed neutrino mass scale.
The nature of the heavy neutrinos is mysterious;
being gauge singlets, their detection in colliders is virtually impossible,
nevertheless they must be present for the seesaw mechanism and leptogenesis.
If our scenario turns out to be correct, CMB would provide a new window to the physics of
right-handed neutrinos.






{\em Acknowledgements.}---
S.K. acknowledges helpful conversation with Kari Enqvist.
This work was supported in part by the Research Program MSM6840770029, 
ATLAS-CERN International Cooperation (M.A.),
the National Research Foundation of Korea Grant No.
2012-007575 (S.K.)
and by the DOE Grant No. DE-FG02-10ER41714 (N.O.).

\bigskip




\end{document}